%% file: EWK-11-004_temp.tex
\pdfoutput=1

\documentclass[11pt,twoside,a4paper,cmspaper,final,collab]{cms-tdr}
\def\svnVersion{266786}\def\cmsCernNo{2013-037}\def\cmsCernDate{\today}\def\cmsMessage{Published in Physics Letters B as \href{http://dx.doi.org/10.1016/j.physletb.2012.10.082}{\doi{10.1016/j.physletb.2012.10.082.}}}

\begin{document}\cmsNoteHeader{EWK-11-004}

\hyphenation{had-ron-i-za-tion}
\hyphenation{cal-or-i-me-ter}
\hyphenation{de-vices}

\RCS$Revision: 156817 $
\RCS$HeadURL: svn+ssh://svn.cern.ch/reps/tdr2/papers/EWK-11-004/trunk/EWK-11-004.tex $
\RCS$Id: EWK-11-004.tex 156817 2012-11-07 08:55:31Z efe $
\newlength\cmsFigWid\setlength{\cmsFigWid}{0.8\textwidth}
\cmsNoteHeader{EWK-11-004} 
\title{Forward-backward asymmetry of Drell--Yan lepton pairs in pp collisions at $\sqrt{s} = $ 7\TeV}

\date{\today}

\abstract{
A measurement of the forward-backward asymmetry ($A_{\mathrm{FB}}$) of Drell--Yan lepton pairs in pp collisions at $\sqrt{s}$ = 7\TeV is presented.
The data sample, collected with the CMS detector, corresponds to
an integrated luminosity of 5\fbinv.
The asymmetry is measured as a function of dilepton mass and rapidity in the
dielectron and dimuon channels.
Combined results from the two channels are presented, and are compared with the standard model predictions.
The $A_{\mathrm{FB}}$ measurement in the dimuon channel and the combination of the two channels are the first such results obtained at a hadron collider.
The measured asymmetries are consistent with the standard model predictions.
}

\hypersetup{%
pdfauthor={CMS Collaboration},%
pdftitle={Forward-backward asymmetry of Drell-Yan lepton pairs in pp collisions at sqrt(s) =  7 TeV},%
pdfsubject={CMS},%
pdfkeywords={CMS, physics, electroweak, Drell-Yan}}

\maketitle 

The amplitude for the standard model (SM) Drell--Yan process
$\cPq\cPaq \rightarrow \cPZ/\gamma^* \rightarrow \ell^+\ell^-$
contains both the vector and the axial-vector couplings of electroweak bosons to fermions~\cite{PhysRevLett.25.316,Drell1971578}.
The differential cross section can be written as
\begin{linenomath}
\begin{equation}
\label{eq:crosssection}
\frac{d\sigma}{d\cos\theta^*} = C\left[\frac{3}{8} (1 + \cos^{2} \theta^*) + A_\mathrm{FB}  \cos\theta^*\right]
\end{equation}
\end{linenomath}
for a given dilepton invariant mass, at leading order, where $\theta^*$ is the emission angle of the lepton ($\ell^-$) relative to the quark momentum in the dilepton centre-of-mass frame.
Forward and backward events are defined by $\cos\theta^*>0$ and $<0$, respectively,
and the asymmetry parameter $A_\mathrm{FB}$ is defined as
\begin{linenomath}
\begin{equation}
A_\mathrm{FB} = \frac{\sigma_{F}-\sigma_{B}}   {\sigma_{F}+\sigma_{B}},
\end{equation}
\end{linenomath}
where $\sigma_{F}$ and $\sigma_{B}$ are the total cross sections for forward and backward events.
Within the SM, the parameters $C$ and $A_\mathrm{FB}$ depend on the vector and axial-vector couplings of the quarks and leptons to
the \cPZ\ boson and on the electric charge of the fermions.

The Drell--Yan cross section is modified by higher-order quantum chromodynamic (QCD) and radiative electroweak corrections.
The electroweak corrections are negligible except near the \cPZ\ peak.
At dilepton masses near the \cPZ\ peak, $A_\mathrm{FB}$ is predicted to be small because of the small value of the lepton vector coupling in the SM, and is sensitive to the electroweak mixing parameter $\sin^2\theta_{\PW}$.
Our measurement of $\sin^2\theta_{\PW}$ with a maximum-likelihood
fit technique based on a smaller data set was reported in Ref.~\cite{PhysRevD.84.112002}.
Above and below the \cPZ\ peak, $A_\mathrm{FB}$ exhibits a characteristic energy dependence governed by virtual photon and \cPZ\ interference.
Deviations from the SM prediction for $A_\mathrm{FB}$ may indicate the existence of particles beyond the standard model~\cite{London:1986, Rosner:1987a, Cvetic:1995, Rosner:1996, Bodek:2001, Abe:1997, Abe:1997b, Davoudiasl:2000}.
If a resonant state exists at high mass, it will interfere with the SM amplitudes and will cause the $A_\mathrm{FB}$ to have a structure near the mass of the new state. Therefore,
studying $A_\mathrm{FB}$ at high mass is particularly useful in a search for resonances that might be missed by a search using the dilepton mass spectrum alone.

The measurement presented in Ref.~\cite{PhysRevD.84.112002} and the $A_\mathrm{FB}$ measurement are complementary. The electroweak mixing parameter was measured within the framework of the SM using events in a dimuon mass window of $M(\mu^+\mu^-) = 80$--100\GeV, while in the current analysis we test the SM and look for signs of new physics at high dilepton mass. The electroweak mixing angle measurement was performed in the dimuon channel. Here we present the results of a measurement of $A_\mathrm{FB}$ with more data, the addition of the dielectron channel, and the combination of the two channels, as a function of mass in a wide mass range and in separate rapidity bins.

To study the forward-backward asymmetry, we use the Collins--Soper frame~\cite{Collins:1977},
in which $\theta^{*}_\mathrm{CS}$
is defined to be the angle between the lepton momentum
and the axis that bisects the angle between the direction of one proton  and
the direction opposite to the other proton in the
centre-of-mass frame of the dilepton.
Use of this frame reduces the uncertainties due to the unknown
transverse momentum of the incoming quarks.
The sign of the longitudinal boost of the dilepton system is used to define
the orientation of the Collins--Soper frame.
The angle $\theta^{*}_\mathrm{CS}$ is calculated from quantities measured in the lab frame as
\begin{linenomath}
\begin{equation}
\label{cos}
\cos\theta^{*}_\mathrm{CS}=\frac{Q_z}{|Q_z|}\frac{2(P^+_1P^-_2-P^-_1P^+_2)}{|Q|\sqrt{Q^2+Q_T^2}},
\end{equation}
\end{linenomath}
where $Q$ is the four-momentum of the dilepton and $Q_{T}$ and $Q_{z}$ are the transverse and longitudinal components of the dilepton momentum with respect to the beam axis; $P_1$ ($P_2$) represents the four-momentum of the lepton (antilepton); and $P_i^\pm=(P_i^0\pm P_i^3)/\sqrt{2}$.
The quark direction is not determined a priori at the Large Hadron Collider (LHC)~\cite{LHC} because
both beams consist of protons.
However, because the antiquark is necessarily a sea quark, on average
we expect it  to carry less momentum than the valence quark, and therefore the dilepton
system is usually boosted in the direction of the valence quark~\cite{Rosner:1987a,Fisher:1994pw,Dittmar:1997}.
This assumption is taken into account by including the sign of the longitudinal boost in the definition of $\cos\theta^{*}_\mathrm{CS}$.
The forward-backward asymmetry is dependent on the dilepton rapidity, $y = \frac{1}{2} \ln [(E + Q_z)/(E - Q_z)]$,
where $E$ and $Q_z$ refer to the energy and the third component of the momentum of the dilepton, respectively.

The raw  $A_\mathrm{FB}$ measurement is distorted compared to the parton-level asymmetry, mainly because of
the dilepton mass resolution of the detector and final-state electromagnetic radiation (FSR).
The asymmetry is further distorted by the detector acceptance and diluted by the imperfect knowledge of the  quark direction at the LHC.
In this Letter we present the $A_\mathrm{FB}$ measurements unfolded to the electroweak vertex (Born level), taking into account the FSR, mass resolution, and other detector effects.
The results are not corrected for the dilution effects due to the acceptance and unknown quark direction because
such corrections require information that is not directly observable.

This analysis is based on a data sample of 5\fbinv collected with the Compact Muon Solenoid (CMS) detector in 2011 at a centre-of-mass energy of 7\TeV.
A detailed description of the CMS detector can be found in~\cite{CMS:2010}.
The central feature of the CMS detector is a 3.8\unit{T} superconducting solenoid of 6\unit{m} internal diameter.
The silicon pixel and strip tracker, the crystal electromagnetic calorimeter
(ECAL), and the brass/scintillator hadron calorimeter are located inside this solenoid.
Muons are measured in the pseudorapidity window $|\eta|<2.4$
using the tracker and the muon system, which is instrumented with detection planes of
three complementary technologies embedded in the steel return yoke of the magnet:
drift tube chambers (DT), cathode strip chambers (CSC), and resistive plate chambers (RPC)~\cite{CMSMUONS:1997}.
Pseudorapidity is defined as $\eta = -\ln[\tan(\theta/2)]$,
where the polar angle $\theta$ is measured with respect to the
anticlockwise-beam direction.
The DT technology is used in the barrel ($|\eta|<1.2$), and CSC in the endcaps
($0.9<|\eta|<2.4$). These are complemented by an RPC system that covers both regions up to $|\eta|<1.6$.
Electrons are detected as energy clusters in the ECAL and as tracks
in the silicon tracker. The ECAL consists of nearly 76\,000 lead tungstate crystals, which provide
coverage in pseudorapidity
$\vert \eta \vert< 1.5 $ in the barrel region and $1.5 <\vert \eta \vert < 3.0$
in the two endcap regions.

The signal ($\cPZ/\gamma^*\rightarrow\Pgmp\Pgmm$,\;\Pep\Pem) and the $\cPZ\rightarrow\Pgt\Pgt$ process, which is considered as a background in this analysis,
are simulated using \POWHEG \cite{Frixione:2007,Alioli:2008,Alioli:2010xd}
at next-to-leading order (NLO).
Parton showering is simulated using \PYTHIA v6.4.24~\cite{Sjostrand:2006} with tune Z2, while
the NLO parton distribution function (PDF) is CT10~\cite{PhysRevD.82.074024}.
The \PW+jets and $\cPqt\cPaqt$ background events are generated using \MADGRAPH~\cite{madgraph5} and
\PYTHIA; the \TAUOLA package is used to describe $\Pgt$ decays~\cite{Davidson:2010rw}.
Event samples of \PW\PW, \PW\cPZ, \cPZ\cPZ, and QCD multijet backgrounds are generated using \PYTHIA.
The generated events are processed with the \GEANTfour-based~\cite{Sulkimo:2003,Allison:2006} CMS detector simulation and reconstructed with the
same software as the collision data.
The signal MC samples include pileup conditions (multiple \Pp\Pp\ interactions occurring in the same bunch crossing) matching those observed in the 2011 data sample.

For data taken in the earlier part of 2011, the dimuon analysis is based on triggers that select events containing at least two muons, each with transverse momentum  $\pt$ of at least 6 or at least 7\GeV, depending on the running period. For the later running period, the triggers select events containing two muons, one with $\pt > 13$\GeV or 17\GeV and the other with $\pt > 8$\GeV.
Within a CSC or DT muon chamber, the hits in the multiple detection layers are fitted to a straight line
representing a segment of the muon track.
In the offline analysis, tracks reconstructed from hits in the silicon tracker are matched to tracks reconstructed from muon segments alone, and then the individual hits in the tracker and muon detectors are refitted to an overall track.
In addition, to increase the acceptance for low momentum muons that may not penetrate deeply into the muon system, tracks from the silicon tracker are extrapolated into the muon system and any that match at least one muon chamber track segment are taken to be muon candidates.
In both cases, multiple scattering and energy loss are taken into account as muons traverse the CMS detector.
Well-reconstructed muons are selected by requiring (1) at least 10 hits in the tracker, including at least one in the pixel detector; (2) at least one segment in the muon system; (3) a normalized $\chi^{2}< 10$ for the overall muon fit (if used); and (4) a transverse distance of closest approach to the beam axis of less than 2~mm.
Cosmic ray muons that traverse CMS close to the interaction point can appear as back-to-back dimuons,
but these are removed by requiring the muon pairs
to have an acollinearity greater than 2.5\unit{mrad}.
Each muon is required to be isolated from other charged tracks based on tracker information alone.
No attempt is made to use radiated photons detected in ECAL to correct muon energies for FSR.
The unfolding procedure corrects for the effect of FSR on $A_\mathrm{FB}$ on a statistical basis.
More details on muon reconstruction and identification can be found in Ref.~\cite{CMSMUO}.
Trigger efficiency factors are calculated and applied for different data-taking periods.
Each muon is required to be within the acceptance of the muon system ($|\eta|\leq2.4$) and have $\pt>20\GeV$.
Events are selected in which opposite-charge muon pairs meet the above requirements.

Dielectron candidates are selected online by requiring two ECAL clusters, each with transverse energy $\ET$ exceeding a threshold value.
Offline reconstruction of electrons starts by building superclusters in the ECAL in order to collect
the energy radiated by bremsstrahlung in the tracker material, following the procedure
described in Ref.~\cite{Baffioni:2006cd}.
A specialized tracking algorithm is used to accommodate changes
of curvature due to bremsstrahlung.
Superclusters are then matched to electron tracks.
Electron candidates are required to have a minimum supercluster $\ET$ of 20\GeV after ECAL energy-scale corrections.
Electrons are restricted to the same phase space as the muons, defined by $\pt>20$\GeV and $|\eta|<2.4$,
for an unambiguous comparison and combination of the two channels.
In order to avoid the inhomogeneous response at the interfaces between the ECAL barrel and endcaps, electrons are further required to fall within
the pseudorapidity ranges $|\eta|\leq1.44$ or $1.57<|\eta|<2.40$.
Electrons are identified by means of shower shape variables, and electron isolation criteria are based on a variable that combines the tracker and calorimeter measurements.
Electrons arising from photon conversions are suppressed by requiring that there be no missing tracker hits before the first hit on the reconstructed track matched to the electron, and also
by rejecting a candidate if it forms a pair with a nearby track that is consistent with a conversion.
More details on electron reconstruction and identification can be found in Ref.~\cite{CMSEGM}.
Energy scale, resolution, and efficiency factors are calculated
and applied for different data-taking periods. Energy scale and resolution factors are derived using $\chi^2$ tests, taking the MC dielectron mass distribution as a constraint.
Events are selected in which opposite-charge electron pairs meet the above requirements.

For both lepton channels, the main sources of background are $\cPZ\rightarrow \Pgt\Pgt$ and QCD dijets for the low mass region and $\cPqt\cPaqt$ for the high mass region.
Diboson (\PW\PW, \PW\cPZ, and \cPZ\cPZ) and inclusive \PW~production processes are lesser sources of background.
Because some QCD jets can pass the electron identification criteria, the QCD background
contribution is non-negligible in the dielectron channel below the \cPZ\ peak.
Electroweak backgrounds are estimated using MC samples.
For both channels, QCD background is estimated from the data under the assumption that same-sign and opposite-sign  lepton pairs are equally probable because the
misidentification of a charged particle in a jet as a lepton or antilepton is equally likely.
Backgrounds are estimated for forward and backward events separately and subtracted bin by bin.
The total background contribution to the data ranges from 0.17\% to 0.21\% in the dimuon channel and from 0.68\% to 0.80\% in the dielectron channel.
After background subtraction, the numbers of events found in the muon channel in the forward and backward regions are
950\,570 and 929\,737, respectively. The corresponding numbers in the electron channel are
448\,338 and 438\,035.

All results are given in the phase-space region defined by
$\pt(\ell)>20\GeV$ and $|\eta(\ell)|<2.4$.
We calculate the $\cos\theta^{*}_\mathrm{CS}$ distributions
in ten bins of dilepton mass $M$ and four bins of rapidity $|y|$,
the limits of which are defined to be
$M = 40$, 50, 60, 76, 86, 96, 106, 120, 150, 200, and 2000\GeV
and $|y| = 0$, 1.00, 1.25, 1.50, and 2.40.

The forward-backward asymmetry is diluted by the events in which
the assumed quark and antiquark directions are incorrect.
The asymmetry is further reduced by the acceptance requirements.
No corrections are applied for either of these effects.
The $|y|<1$ bin has the largest asymmetry dilution
due to the unknown quark direction,
but the smallest acceptance effect.
The next two bins, $1.00<|y|<1.25$ and $1.25<|y|<1.50$,
have the largest asymmetry.
The highest rapidity bin, $1.50<|y|<2.40$,
is least affected by the unknown quark direction
but suffers a large acceptance reduction
resulting in a smaller asymmetry compared to other $|y|$ bins.

To correct for FSR, mass resolution, efficiencies, and other detector effects,
we unfold the forward and backward mass spectra in each $|y|$ bin.
The unfolding procedure is performed using a matrix inversion
technique~\cite{Blobel:2002pu}.
The unfolding is performed with response matrices that provide
a mapping between the corrected and measured numbers of events
in each mass and rapidity bin:
\begin{linenomath}
\begin{equation}
\left.N_{j}^\text{meas}(F,k) = \sum_{i=1}^{10}R_{ji}^{FF}(k)\,N_i^\text{corrected}(F,k)\;\right[\begin{aligned}j&=1,\ldots,10\\k&=1,\ldots,4\end{aligned}
\end{equation}
\end{linenomath}
\begin{linenomath}
\begin{equation}
\left.N_{j}^\text{meas}(B,k) = \sum_{i=1}^{10}R_{ji}^{BB}(k)\,N_i^\text{corrected}(B,k)\;\right[\begin{aligned}j&=1,\ldots,10\\k&=1,\ldots,4\end{aligned}.
\end{equation}
\end{linenomath}
In these equations, $N_j(F,k)$ and $N_j(B,k)$ refer to the number of forward ($F$) and backward ($B$) events within the acceptance ($p_T>20$\GeV and $|\eta|<2.4$) in each mass bin~$j$ for the rapidity bin~$k$;
$R^{FF}_{ji}(k)$ is the response matrix describing the transfer of forward events from generated mass bin~$i$ to observed mass bin~$j$, while $R^{BB}_{ji}(k)$ is the response matrix for the backward events.
We construct the response matrices for unfolding the reconstructed forward and backward mass spectra in each $|y|$ bin to the Born level.
The response matrices are calculated using MC events before and after simulation of FSR and the detector effects. Therefore they account for the FSR and mass resolution as well as the efficiency within the detector acceptance. 
In the dielectron channel, the gap in ECAL in the pseudorapidity range of $1.44<|\eta|<1.57$ is
treated as an inefficiency and corrected by the unfolding procedure.
The response matrices that represent the forward generated but backward reconstructed events (and vice versa) have a negligible contribution and are not used in this study, but the effect of this approximation is taken into account in the systematic uncertainties.
The unfolded values are obtained by inverting the above equations.  The corresponding
uncertainties are calculated taking into account the correlations due to the unfolding procedure.
The estimated uncertainties are verified by applying the procedure to a large number of independent MC samples.
These MC samples are also used to check whether there is a bias in the $A_\mathrm{FB}$ values obtained through unfolding,
and the maximum difference in $A_\mathrm{FB}$ is found to be 0.06.

Systematic uncertainties are estimated in each $M$-$|y|$ bin using MC events.
All systematic uncertainties are assumed to be independent and are combined in quadrature.

Although the background is small in the Drell--Yan process, uncertainties in the background estimation lead to systematic errors in the final results. We take a conservative approach and assume that this small background is uncertain by 100\%, and therefore scale the background up and down by 100\% and repeat the analysis.
The largest difference from the nominal $A_\mathrm{FB}$ is found to be $0.04$. The systematic uncertainty
in the background estimate in all other bins is smaller than $0.03$.

To quantify possible systematic uncertainties that could arise from the modelling of FSR in \PYTHIA, we examine the events that show the largest change in lepton momentum pre- and post-radiation.
The \PYTHIA description of FSR agrees with data to within $\pm$5\%, so we reweight by $\pm$5\%
the events  for which the difference of the momenta of a lepton pre- and post-radiation is larger than
1\GeV.
The distributions obtained from the reweighted events are used to obtain new values for $A_\mathrm{FB}$.
Even such a large change in event weights results in a change in the value of $A_\mathrm{FB}$ of less than
0.02.
These changes in the value of $A_\mathrm{FB}$ are assigned as systematic uncertainties arising from uncertainty in the modelling of FSR.

The systematic uncertainties related to detector alignment
are studied using MC samples with
different assumed tracker reconstruction geometries (basic distortions) based on the cylindrical symmetry of the tracker system~\cite{Chatrchyan:2009sr}.
The differences between the $A_\mathrm{FB}$ values obtained with the ideal geometry and the other scenarios are evaluated.
For each $M$-$|y|$ bin, the largest difference is taken as the alignment uncertainty.
The largest of these uncertainties over the entire $M$-$|y|$ range is 0.01.

In the dielectron channel, the uncertainties obtained from the $\chi^2$ minimization used to obtain the energy-scale and resolution factors are used to modify the energy scale and hence calculate the associated uncertainties.
In the dimuon channel, no energy scale or resolution factors are applied, but to account for a possible scale uncertainty the energy scale is changed by 0.1\% and the analysis is repeated. The resulting mass shift is found to be negligible.
The largest systematic uncertainty due to energy scale and resolution is found to be 0.02 in $A_\mathrm{FB}$.

The trigger efficiency uncertainties are estimated by comparing the results before and after
trigger scale factors are applied for both channels.
The uncertainties due to pileup are estimated by comparing results with different pileup multiplicity profiles for both channels.
The efficiency uncertainties are found to be smaller than 0.005 and the pileup reweighting uncertainties smaller than 0.03 in $A_\mathrm{FB}$.

The resulting total experimental systematic uncertainty is at most 0.1 in $A_\mathrm{FB}$; however, for most of the bins the total experimental uncertainty is less than 0.05.

The total experimental systematic uncertainty does not include the PDF or $\alpha_s$ uncertainties.
To determine these, we follow the recommendation of the PDF4LHC working group~\cite{Alekhin:2011sk,Botje:2011sn}.
At the NLO level, the recommendation is to reweight a sample generated with the CT10 PDF set~\cite{PhysRevD.78.013004,PhysRevD.82.074024}
to obtain samples that mimic the NNPDF2.1~\cite{Ball:2010de} and MSTW2008~\cite{Martin:2009iq} PDFs.
The internal degrees of freedom of each PDF set are varied.
Samples corresponding to different $\alpha_s(M_Z)$ assumptions are obtained in a similar manner.
The value of $A_\mathrm{FB}$ is calculated in each $M$-$|y|$ bin for each variation.
The resulting variations in $A_\mathrm{FB}$ are combined to
obtain the PDF uncertainty, following the PDF4LHC prescription.
The largest uncertainty is found to be 0.012.

\begin{figure*}[Ht!bp]
\begin{center}
\includegraphics[width=\cmsFigWid]{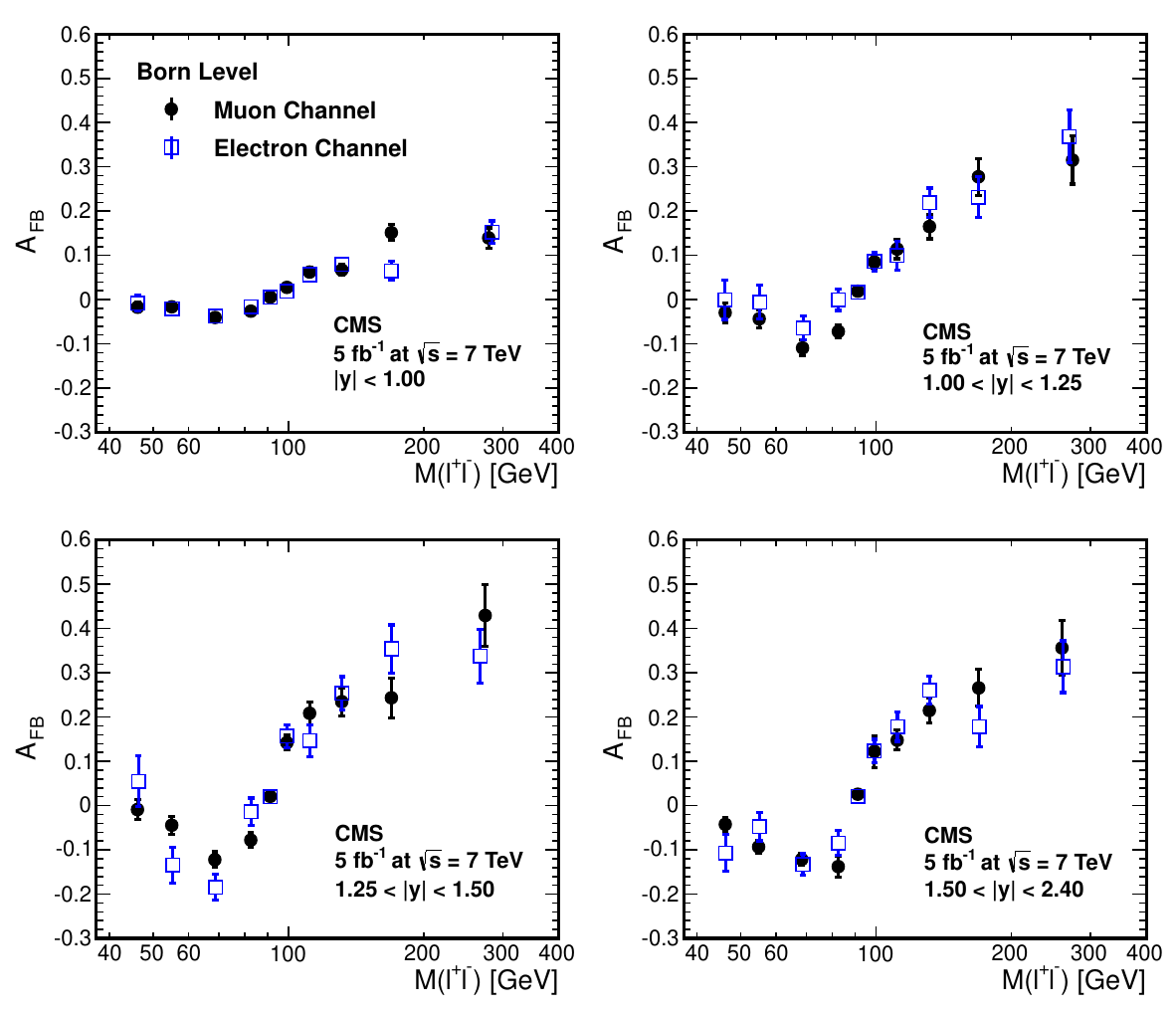}
\caption{The unfolded $\Pgmp\Pgmm$ and $\Pep\Pem$  measurements of $A_\mathrm{FB}$ at the Born level
in four $|y|$ bins for $\pt(\ell) > 20\GeV$ and $|\eta(\ell)| < 2.4$. The data points are shown with statistical error bars.}
\label{fig:AFB_electron_muon}
\end{center}
\end{figure*}

\begin{figure*}[htbp!]
\begin{center}
\includegraphics[width=\cmsFigWid]{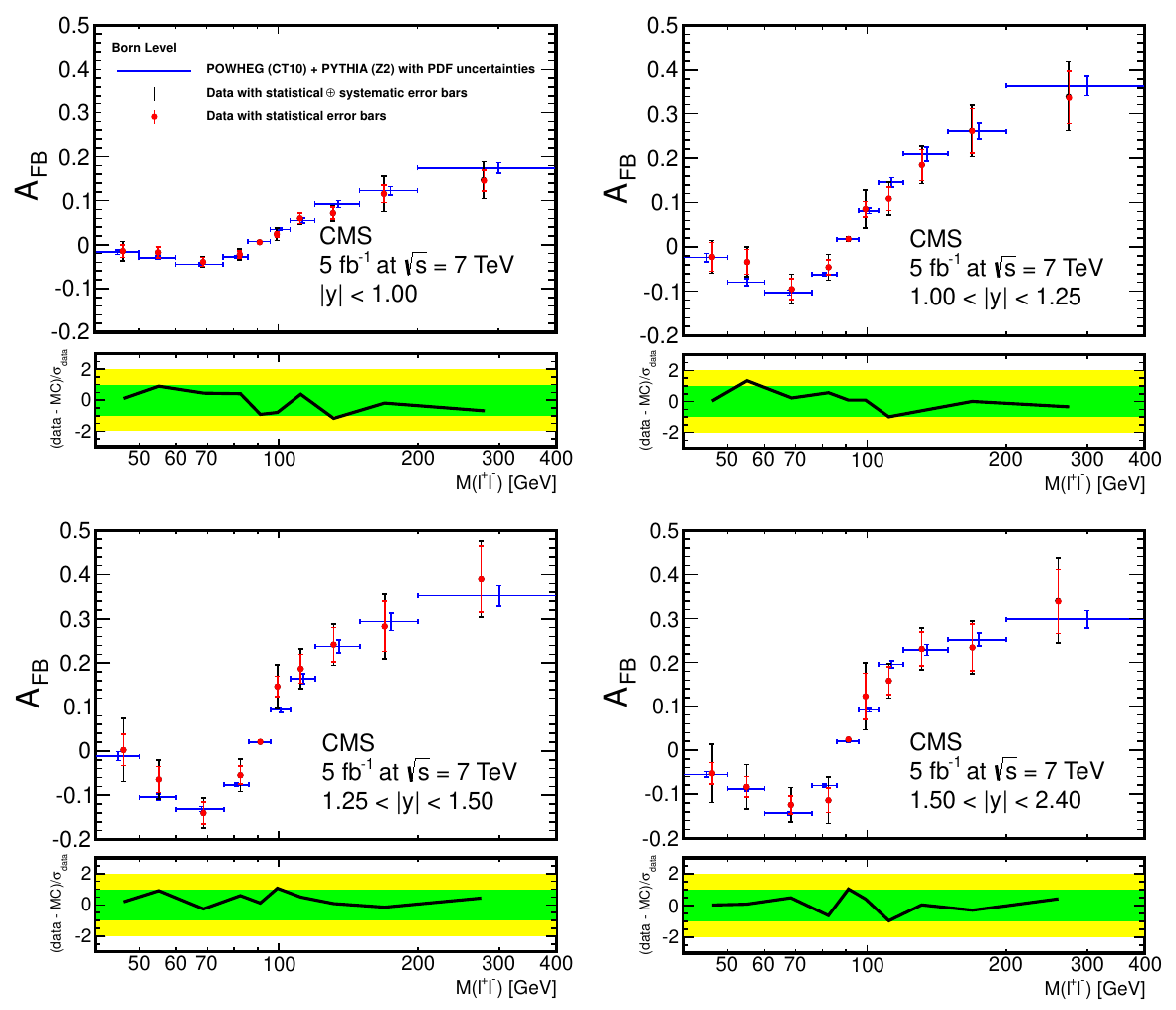}
\caption{The unfolded and combined ($\Pgmp\Pgmm$ and $\Pep\Pem$) measurement of $A_\mathrm{FB}$ at the Born level
in four $|y|$ bins for $\pt(\ell) > 20\GeV$ and $|\eta(\ell)| < 2.4$.
The data points are shown with both statistical error bars and combined statistical and systematic error bars.
The error bars on the MC points represent the quadratically summed PDF uncertainties and statistical errors.
The horizontal extent of the error bars indicates the bin width (except for the last bin, which is truncated at 400\GeV).
Beneath each plot is shown the difference between data and
MC, normalized by the combined statistical and systematic uncertainty.
The green and yellow bands indicate the $1\sigma$ and $2\sigma$ differences of data from theory predictions.
}
\label{fig:AFB_combined_Born}
\end{center}
\end{figure*}

The unfolded AFB distributions at the Born level for the dimuon and
dielectron channels are shown in Fig.~\ref{fig:AFB_electron_muon}.
The measurements in the two channels agree well with each other.
The unfolded and combined $A_\mathrm{FB}$ distributions at the Born level
are displayed in Fig.~\ref{fig:AFB_combined_Born} and Table~\ref{tab:table1}.
All these distributions are in agreement with the SM expectations and
there is no indication of non-SM physics.
The reversal of the sign of $A_\mathrm{FB}$ near the \cPZ\ peak is due to the change
of sign of the \cPZ-$\gamma^*$ interference term.
The asymmetry is already evident at the raw level, before
unfolding, and the bins that are most affected by unfolding are
those just below and just above the \cPZ\  peak.
Table~\ref{tab:table1} also shows the difference of the unfolded and raw asymmetries
in each bin.
Table~\ref{tab:table2}  shows the estimated systematic uncertainties in each rapidity
bin for the mass bin around the Z peak.

\begin{table*}[htbp]
\centering
\topcaption{Unfolded combined measurements of $A_{\mathrm{FB}}$ in each $M$-$|y|$ bin. The average mass in each bin is shown, together with the measured $A_{\mathrm{FB}}$ and the corresponding statistical, systematic, and total uncertainties.
The statistical and systematic errors are combined in quadrature to obtain the total uncertainties.
The $A_{\mathrm{FB}}$ values estimated from MC are also shown, with the corresponding statistical and PDF uncertainties.
The final column shows the larger of the differences $A_{\text{FB}}({\text{Born}}) - A_{\text{FB}}({\text{raw}})$
in the muon and electron channels.
 }
\ifthenelse{\boolean{cms@external}}{\footnotesize}{\tiny}

\begin{tabular}{cD{-}{\mbox{--}}{-1}D{.}{.}{3.1}D{.}{.}{2.3}D{.}{.}{1.3}D{.}{.}{1.3}D{.}{.}{1.3}D{.}{.}{2.3}D{.}{.}{1.3}D{.}{.}{1.3}D{.}{.}{2.3}}
\hline
  \multicolumn{1}{c}{\abs{y}} & \multicolumn{1}{c}{M [\GeVns{}]} &  \multicolumn{1}{c}{$\langle M \rangle$ [\GeVns{}]} &   \multicolumn{1}{c}{$A_{\mathrm{FB}}$(data)} & \multicolumn{1}{c}{stat. err.} &  \multicolumn{1}{c}{syst. err.} &  \multicolumn{1}{c}{ tot. err.} & \multicolumn{1}{c}{$A_{\mathrm{FB}}$(MC)}  &   \multicolumn{1}{c}{stat. err. (MC)}
& \multicolumn{1}{c}{PDF err.} & \multicolumn{1}{c}{Unfolding}\\
\hline
      & 40-50 & 46.2 & -0.014  & 0.014  &   0.017  &  0.022  &   -0.017 & 0.005 & 0.001 & -0.002\\
      & 50-60 & 55.0 & -0.017  & 0.012  &   0.005  &  0.013  &   -0.029 & 0.004 & 0.001  & 0.001\\
      & 60-76 & 68.7 & -0.039  & 0.010  &   0.005  &  0.011  &   -0.044 & 0.003 & 0.002  & -0.015\\
      & 76-86 & 82.5 & -0.022  & 0.008  &   0.010  &  0.013  &   -0.028 & 0.002 & 0.001  & -0.015\\
 0--1 & 86-96 & 91.1 & 0.006  & 0.001  &   0.001  &  0.002  &   0.007   & 0.001 & 0.001   &-0.001\\
 &           96-106 & 99.3 & 0.024  & 0.006  &   0.012  &  0.014  &   0.035 & 0.002 & 0.002   & 0.003\\
 &           106-120 & 111.5 & 0.060  & 0.012  &   0.006  &  0.013  &   0.055 & 0.004 & 0.004 & 0.003\\
 &           120-150 & 131.5 & 0.072  & 0.014  &   0.010  &  0.017  &   0.092 & 0.006 & 0.006 & -0.007\\
 &           150-200 & 169.5 & 0.116  & 0.020  &   0.034  &  0.040  &   0.123 & 0.009 & 0.006 & 0.011\\
 &           200-400 & 278.6 & 0.147  & 0.024  &   0.035  &  0.042  &   0.174 & 0.011 & 0.007 & 0.001\\
\hline
 &         40-50 & 46.2 & -0.022  & 0.032  &   0.019  &  0.037  &   -0.024 & 0.011 & 0.001    & -0.004\\
 &           50-60 & 55.0 & -0.034  & 0.028  &   0.019  &  0.034  &   -0.079 & 0.008 & 0.002  & -0.005\\
 &           60-76 & 68.7 & -0.095  & 0.023  &   0.025  &  0.034  &   -0.103 & 0.005 & 0.001  & -0.045\\
 &           76-86 & 82.5 & -0.046  & 0.017  &   0.024  &  0.029  &   -0.062 & 0.003 & 0.001  & -0.051\\
 1--1.25 & 86-96 & 91.1 & 0.018  & 0.003  &   0.004  &  0.005  &   0.018 & 0.001 & 0.001 & -0.001\\
 &           96-106 & 99.2 & 0.085  & 0.017  &   0.039  &  0.043  &   0.081 & 0.004 & 0.004   & -0.021\\
 &           106-120 & 111.5 & 0.109  & 0.026  &   0.026  &  0.037  &   0.146 & 0.009 & 0.007 &  0.002\\
 &           120-150 & 131.6 & 0.185  & 0.035  &   0.024  &  0.042  &   0.210 & 0.012 & 0.012 & -0.002\\
 &           150-200 & 169.0 & 0.261  & 0.050  &   0.029  &  0.058  &   0.260 & 0.018 & 0.009 &  0.006\\
 &           200-400 & 273.1 & 0.340  & 0.060  &   0.051  &  0.078  &   0.364 & 0.024 & 0.010 &  0.005\\
\hline
 &          40-50 & 46.2 & 0.002  & 0.035  &   0.062  &  0.072  &   -0.012 & 0.011& 0.002    & -0.002\\
 &           50-60 & 55.0 & -0.064  & 0.030  &   0.033  &  0.044  &   -0.104 & 0.009& 0.002  & -0.004\\
 &           60-76 & 68.6 & -0.140  & 0.025  &   0.023  &  0.034  &   -0.131 & 0.005& 0.001  & -0.055\\
 &           76-86 & 82.5 & -0.055  & 0.020  &   0.030  &  0.036  &   -0.076 & 0.003& 0.001  & -0.056\\
 1.25--1.5 &86-96 & 91.1 & 0.021  & 0.003  &   0.003  &  0.004  &   0.020 & 0.001& 0.001 & -0.001\\
 &           96-106 & 99.3 & 0.147  & 0.023  &   0.044  &  0.050  &   0.094 & 0.004& 0.005   & 0.050\\
 &           106-120 & 111.5 & 0.187  & 0.032  &   0.031  &  0.045  &   0.164 & 0.010& 0.007 & 0.005\\
 &           120-150 & 131.5 & 0.242  & 0.039  &   0.027  &  0.047  &   0.238 & 0.013& 0.010 & 0.013\\
 &           150-200 & 169.6 & 0.283  & 0.057  &   0.046  &  0.073  &   0.294 & 0.019& 0.012 & -0.005\\
 &           200-400 & 274.3 & 0.390  & 0.074  &   0.043  &  0.086  &   0.352 & 0.026& 0.010 &  0.004\\
\hline
 &          40-50 & 46.2 & -0.053  & 0.024  &   0.061  &  0.066  &     -0.055 & 0.007& 0.001 & -0.006 \\
 &           50-60 & 54.8 & -0.084  & 0.023  &   0.044  &  0.050  &    -0.088 & 0.006& 0.001 & -0.006\\
 &           60-76 & 68.4 & -0.124  & 0.020  &   0.033  &  0.039  &    -0.143 & 0.004& 0.001 & -0.053\\
 &           76-86 & 82.5 & -0.114  & 0.028  &   0.045  &  0.053  &    -0.080 & 0.002& 0.001 & -0.127 \\
 1.5--2.4&86-96 & 91.1 & 0.024  & 0.003  &   0.003  &   0.004  &     0.020 & 0.001& 0.001 &  0.002 \\
 &           96-106 & 99.3 & 0.123  & 0.053  &   0.055  &  0.076  &     0.091 & 0.003& 0.003 &  0.059\\
 &           106-120 & 111.5 & 0.158  & 0.031  &   0.024  &  0.039  &   0.196 & 0.007& 0.005 &  -0.003\\
 &           120-150 & 131.5 & 0.231  & 0.038  &   0.028  &  0.047  &   0.229 & 0.010& 0.009 &  0.004\\
 &           150-200 & 169.4 & 0.234  & 0.053  &   0.028  &  0.060  &   0.252 & 0.015& 0.007 &  0.009\\
 &           200-400 & 259.1 & 0.340  & 0.072  &   0.063  &  0.096  &   0.298 & 0.022& 0.006 &  0.028\\
\hline
\end{tabular}
\label{tab:table1}
\end{table*}

\begin{table*}[htbp]
\centering
\topcaption{Estimated systematic uncertainties on $A_{\mathrm{FB}}$, in units of $10^{-3}$, for each rapidity bin, in the mass bin
around the \cPZ\ peak, $M = 86$--96\GeV. The components are discussed in the text.}
\begin{tabular}{cllll}
\hline
$|y|$ & \multicolumn{1}{c}{0--1} & \multicolumn{1}{c}{1--1.25} & \multicolumn{1}{c}{1.25--1.5} & \multicolumn{1}{c}{1.5--2.4} \\
\hline
FSR & $\pm$0.1  &+0.4/$-0.1$   & +0.8/$-0.1$   & +6.2/$-0.2$   \\
Energy scale &$\pm$0.1    & +0.4/$-0.1$ & +0.9/$-0.5$  &+0.3/$-0.4$    \\
Resolution & +0.1/$-0.2$ & +0.6/$-0.5$ & $\pm$0.2   & +0.0/$-0.9$  \\
Alignment & +0.4/$-0.1$ & +0.5/$-0.1$  & +0.7/$-0.0$   & +0.0/$-1.9$  \\
Background & $\pm$0.1  & $\pm$0.1  & $\pm$0.1  & $\pm$0.1  \\
Pileup and Eff. & +0.2/$-1.2$  & +0.3/$-0.9$ & +1.9/$-0.3$   & +0.5/$-1.1$ \\
Unfolding & +0.1/$-0.0$   & +3.5/$-0.0$ & +1.4/$-0.0$ & +1.1/$-0.0$ \\
PDFs  & $\pm$0.6 & $\pm$0.4 & $\pm$1.4 & $\pm$1.4 \\
\hline

\hline
\end{tabular}
\label{tab:table2}
\end{table*}

In summary, we have presented a measurement of the forward-backward asymmetry $A_\mathrm{FB}$ for opposite-charge lepton pairs produced via an intermediate
Z/$\gamma^*$ at $\sqrt{s}$ = 7\TeV in the CMS experiment, based on a sample of pp collisions corresponding to an integrated luminosity of 5\fbinv.
The asymmetry is studied as a function of the dilepton rapidity and the dilepton mass $M$ for  $M> 40\GeV.$
The unfolded and combined measurements at the Born level are presented.
We find the $A_\mathrm{FB}$ distributions to be consistent with the standard model predictions.
The $A_\mathrm{FB}$ measurement in the dimuon channel and the combination of the two channels are the first such results obtained at a hadron collider.

\section*{Acknowledgements}
We congratulate our colleagues in the CERN accelerator departments for the excellent performance of the LHC machine. We thank the technical and administrative staff at CERN and other CMS institutes, and acknowledge support from: FMSR (Austria); FNRS and FWO (Belgium); CNPq, CAPES, FAPERJ, and FAPESP (Brazil); MES (Bulgaria); CERN; CAS, MoST, and NSFC (China); COLCIENCIAS (Colombia); MSES (Croatia); RPF (Cyprus); MoER, SF0690030s09 and ERDF (Estonia); Academy of Finland, MEC, and HIP (Finland); CEA and CNRS/IN2P3 (France); BMBF, DFG, and HGF (Germany); GSRT (Greece); OTKA and NKTH (Hungary); DAE and DST (India); IPM (Iran); SFI (Ireland); INFN (Italy); NRF and WCU (Korea); LAS (Lithuania); CINVESTAV, CONACYT, SEP, and UASLP-FAI (Mexico); MSI (New Zealand); PAEC (Pakistan); MSHE and NSC (Poland); FCT (Portugal); JINR (Armenia, Belarus, Georgia, Ukraine, Uzbekistan); MON, RosAtom, RAS and RFBR (Russia); MSTD (Serbia); MICINN and CPAN (Spain); Swiss Funding Agencies (Switzerland); NSC (Taipei); TUBITAK and TAEK (Turkey); STFC (United Kingdom); DOE and NSF (USA).
\bibliography{auto_generated}   
\cleardoublepage \appendix\section{The CMS Collaboration \label{app:collab}}\begin{sloppypar}\hyphenpenalty=5000\widowpenalty=500\clubpenalty=5000\input{EWK-11-004-authorlist.tex}\end{sloppypar}
\end{document}

%% file: EWK-11-004-authorlist.tex
\textbf{Yerevan Physics Institute,  Yerevan,  Armenia}\\*[0pt]
S.~Chatrchyan, V.~Khachatryan, A.M.~Sirunyan, A.~Tumasyan
\vskip\cmsinstskip
\textbf{Institut f\"{u}r Hochenergiephysik der OeAW,  Wien,  Austria}\\*[0pt]
W.~Adam, T.~Bergauer, M.~Dragicevic, J.~Er\"{o}, C.~Fabjan\cmsAuthorMark{1}, M.~Friedl, R.~Fr\"{u}hwirth\cmsAuthorMark{1}, V.M.~Ghete, J.~Hammer, N.~H\"{o}rmann, J.~Hrubec, M.~Jeitler\cmsAuthorMark{1}, W.~Kiesenhofer, V.~Kn\"{u}nz, M.~Krammer\cmsAuthorMark{1}, D.~Liko, I.~Mikulec, M.~Pernicka$^{\textrm{\dag}}$, B.~Rahbaran, C.~Rohringer, H.~Rohringer, R.~Sch\"{o}fbeck, J.~Strauss, A.~Taurok, P.~Wagner, W.~Waltenberger, G.~Walzel, E.~Widl, C.-E.~Wulz\cmsAuthorMark{1}
% ...
\vskip\cmsinstskip
\dag:~Deceased\\
1:~~Also at Vienna University of Technology, Vienna, Austria\\
% ...


%% file: EWK-11-004_temp.bbl
\providecommand{\href}[2]{#2}\begingroup\raggedright\begin{thebibliography}{10}%
\makeatletter
\providecommand{\hrefCMSnoop }[0]{\@secondoftwo}%
\makeatother
\providecommand{\doi}{\texttt{doi:}\begingroup \urlstyle{tt}\Url}

\bibitem{PhysRevLett.25.316}
\hrefCMSnoop {} {S.~D. Drell and T.-M. Yan, ``Massive Lepton-Pair Production in
  Hadron-Hadron Collisions at High Energies'',} \textit{ Phys. Rev. Lett.}
  \textbf{ 25} (1970) 316,
  \href{http://dx.doi.org/10.1103/PhysRevLett.25.316}{\doi{10.1103/PhysRevLett.25.316}}.

\bibitem{Drell1971578}
\hrefCMSnoop {} {S.~D. Drell and T.-M. Yan, ``Partons and their applications at
  high energies'',} \textit{ Annals of Physics} \textbf{ 66} (1971) 578,
  \href{http://dx.doi.org/10.1016/0003-4916(71)90071-6}{\doi{10.1016/0003-4916(71)90071-6}}.

\bibitem{PhysRevD.84.112002}
\hrefCMSnoop {} {{ CMS} Collaboration, ``Measurement of the weak mixing angle
  with the {Drell--Yan} process in proton-proton collisions at the {LHC}'',}
  \textit{ Phys. Rev. D} \textbf{ 84} (2011) 112002,
  \href{http://dx.doi.org/10.1103/PhysRevD.84.112002}{\doi{10.1103/PhysRevD.84.112002}}.

\bibitem{London:1986}
\hrefCMSnoop {} {D.~London and J.~L. Rosner, ``Extra Gauge Bosons in
  {E$_6$}'',} \textit{ Phys. Rev. D} \textbf{ 34} (1986) 1530,
  \href{http://dx.doi.org/10.1103/PhysRevD.34.1530}{\doi{10.1103/PhysRevD.34.1530}}.

\bibitem{Rosner:1987a}
\hrefCMSnoop {} {J.~L. Rosner, ``Off-Peak Lepton Asymmetries from New {Z's}'',}
  \textit{ Phys. Rev. D} \textbf{ 35} (1987) 2244,
  \href{http://dx.doi.org/10.1103/PhysRevD.35.2244}{\doi{10.1103/PhysRevD.35.2244}}.

\bibitem{Cvetic:1995}
\hrefCMSnoop {} {M.~Cveti\v{c} and S.~Godfrey, ``Discovery and Identification
  of Extra Gauge Bosons'',} in \textit{ Electroweak Symmetry Breaking and New
  Physics at the {TeV} Scale}, T.~L. Barklow, S.~Dawson, H.~E. Haber, and J.~L.
  Siegrist, eds.
\newblock World Scientific, 1995.
\newblock
\href{http://www.arXiv.org/abs/hep-ph/9504216}{\texttt{ arXiv:hep-ph/9504216}}.
\newblock

\bibitem{Rosner:1996}
\hrefCMSnoop {} {J.~L. Rosner, ``Forward-Backward Asymmetries in Hadronically
  Produced Lepton Pairs'',} \textit{ Phys. Rev. D} \textbf{ 54} (1996) 1078,
  \href{http://dx.doi.org/10.1103/PhysRevD.54.1078}{\doi{10.1103/PhysRevD.54.1078}}.

\bibitem{Bodek:2001}
\hrefCMSnoop {} {A.~Bodek and U.~Baur, ``Implications of a {300--$500\GeVcc$
  Z'} boson on $p\bar{p}$ collider data at $\sqrt{s} = 1.8$~{TeV}'',} \textit{
  Eur. Phys. J. C} \textbf{ 21} (2001) 607,
  \href{http://dx.doi.org/10.1007/s100520100778}{\doi{10.1007/s100520100778}}.

\bibitem{Abe:1997}
\hrefCMSnoop {} {{ CDF} Collaboration, ``Search for New Gauge Bosons Decaying
  into Dileptons in $\bar{p}p$ Collisions at $\sqrt{s} = 1.8$~{TeV}'',}
  \textit{ Phys. Rev. Lett.} \textbf{ 79} (1997) 2192,
  \href{http://dx.doi.org/10.1103/PhysRevLett.79.2192}{\doi{10.1103/PhysRevLett.79.2192}}.

\bibitem{Abe:1997b}
\hrefCMSnoop {} {{ CDF} Collaboration, ``Limits on Quark-Lepton Compositeness
  Scales from Dileptons Produced in 1.8 {TeV} $p\bar{p}$ Collisions'',}
  \textit{ Phys. Rev. Lett.} \textbf{ 79} (1997) 2198,
  \href{http://dx.doi.org/10.1103/PhysRevLett.79.2198}{\doi{10.1103/PhysRevLett.79.2198}}.

\bibitem{Davoudiasl:2000}
\hrefCMSnoop {} {H.~Davoudiasl, J.~L. Hewett, and T.~G. Rizzo, ``Phenomenology
  of the {Randall--Sundrum} Gauge Hierarchy Model'',} \textit{ Phys. Rev.
  Lett.} \textbf{ 84} (2000) 2080,
  \href{http://dx.doi.org/10.1103/PhysRevLett.84.2080}{\doi{10.1103/PhysRevLett.84.2080}}.

\bibitem{Collins:1977}
\hrefCMSnoop {} {J.~C. Collins and D.~E. Soper, ``Angular Distribution of
  Dileptons in High-Energy Hadron Collisions'',} \textit{ Phys. Rev. D}
  \textbf{ 16} (1977) 2219,
  \href{http://dx.doi.org/10.1103/PhysRevD.16.2219}{\doi{10.1103/PhysRevD.16.2219}}.

\bibitem{LHC}
\hrefCMSnoop {} {L.~Evans and P.~Bryant, ``{LHC} Machine'',} \textit{ JINST}
  \textbf{ 03} (2008) S08001,
  \href{http://dx.doi.org/10.1088/1748-0221/3/08/S08001}{\doi{10.1088/1748-0221/3/08/S08001}}.

\bibitem{Fisher:1994pw}
\hrefCMSnoop {} {P.~Fisher, U.~Becker, and J.~Kirkby, ``{Very high precision
  tests of the electroweak theory}'',} \textit{ Phys. Lett. B} \textbf{ 356}
  (1995) 404,
\href{http://dx.doi.org/10.1016/0370-2693(95)00714-V}{\doi{10.1016/0370-2693(95)00714-V}}.

\bibitem{Dittmar:1997}
\hrefCMSnoop {} {M.~Dittmar, ``Neutral current interference in the {TeV}
  region: The experimental sensitivity at the {CERN LHC}'',} \textit{ Phys.
  Rev. D} \textbf{ 55} (1997) 161,
  \href{http://dx.doi.org/10.1103/PhysRevD.55.161}{\doi{10.1103/PhysRevD.55.161}}.

\bibitem{CMS:2010}
\hrefCMSnoop {} {{ CMS} Collaboration, ``The {CMS} experiment at the {CERN}
  {LHC}'',} \textit{ JINST} \textbf{ 3} (2008) S08004,
\href{http://dx.doi.org/10.1088/1748-0221/3/08/S08004}{\doi{10.1088/1748-0221/3/08/S08004}}.

\bibitem{CMSMUONS:1997}
\href {http://cdsweb.cern.ch/record/343814} {{CMS Collaboration}, ``{CMS} MUON
  Technical Design Report'',} CMS TDR CERN-LHCC-97-032; CMS-TDR-003, 1997.

\bibitem{Frixione:2007}
\hrefCMSnoop {} {S.~Frixione, P.~Nason, and C.~Oleari, ``Matching NLO QCD
  Computations with Parton Shower simulations: the POWHEG method'',} \textit{
  JHEP} \textbf{ 11} (2007) 070,
  \href{http://dx.doi.org/10.1088/1126-6708/2007/11/070}{\doi{10.1088/1126-6708/2007/11/070}}.

\bibitem{Alioli:2008}
\hrefCMSnoop {} {S.~Alioli, P.~Nason, C.~Oleari, and E.~Re, ``NLO Vector-Boson
  Production Matched with Shower in POWHEG'',} \textit{ JHEP} \textbf{ 07}
  (2008) 060,
  \href{http://dx.doi.org/10.1088/1126-6708/2008/07/060}{\doi{10.1088/1126-6708/2008/07/060}}.

\bibitem{Alioli:2010xd}
\hrefCMSnoop {} {S.~Alioli, P.~Nason, C.~Oleari, and E.~Re, ``{A general
  framework for implementing NLO calculations in shower Monte Carlo programs:
  the POWHEG BOX}'',} \textit{ JHEP} \textbf{ 06} (2010) 043,
  \href{http://dx.doi.org/10.1007/JHEP06(2010)043}{\doi{10.1007/JHEP06(2010)043}},
\href{http://www.arXiv.org/abs/1002.2581}{\texttt{ arXiv:1002.2581}}.

\bibitem{Sjostrand:2006}
\hrefCMSnoop {} {T.~Sj{\"o}strand, S.~Mrenna, and P.~Z. Skands, ``{PYTHIA} 6.4
  physics and manual'',} \textit{ JHEP} \textbf{ 05} (2006) 026,
  \href{http://dx.doi.org/10.1088/1126-6708/2006/05/026}{\doi{10.1088/1126-6708/2006/05/026}}.

\bibitem{PhysRevD.82.074024}
H.-L. Lai\hrefCMSnoop {} { {et~al.}, ``New parton distributions for collider
  physics'',} \textit{ Phys. Rev. D} \textbf{ 82} (2010) 074024,
  \href{http://dx.doi.org/10.1103/PhysRevD.82.074024}{\doi{10.1103/PhysRevD.82.074024}}.

\bibitem{madgraph5}
J.~Alwall\hrefCMSnoop {} { {et~al.}, ``{MadGraph 5}: going beyond'',} \textit{
  JHEP} \textbf{ 06} (2011) 128,
  \href{http://dx.doi.org/10.1007/JHEP06(2011)128}{\doi{10.1007/JHEP06(2011)128}},
\href{http://www.arXiv.org/abs/1106.0522}{\texttt{ arXiv:1106.0522}}.

\bibitem{Davidson:2010rw}
N.~Davidson\hrefCMSnoop {} { {et~al.}, ``Universal Interface of TAUOLA
  Technical and Physics Documentation'',} (2010).
\href{http://www.arXiv.org/abs/1002.0543}{\texttt{ arXiv:1002.0543}}.

\bibitem{Sulkimo:2003}
\hrefCMSnoop {} {S.~Agostinelli {et~al.}, ``{Geant4}--a simulation toolkit'',}
  \textit{ Nucl. Instrum. Meth. A} \textbf{ 506} (2003) 250,
  \href{http://dx.doi.org/10.1016/S0168-9002(03)01368-8}{\doi{10.1016/S0168-9002(03)01368-8}}.

\bibitem{Allison:2006}
\hrefCMSnoop {} {J.~Allison {et~al.}, ``Geant4 developments and
  applications'',} \textit{ IEEE Transactions on Nuclear Science} \textbf{ 53}
  (2006) 270,
  \href{http://dx.doi.org/10.1109/TNS.2006.869826}{\doi{10.1109/TNS.2006.869826}}.

\bibitem{CMSMUO}
\href {http://cdsweb.cern.ch/record/1279140} {{ CMS} Collaboration,
  ``Performance of muon identification in pp collisions at $\sqrt{s}$ = 7
  {TeV}'',} CMS Physics Analysis Summary CMS-PAS-MUO-10-002, 2010.

\bibitem{Baffioni:2006cd}
S.~Baffioni\hrefCMSnoop {} { {et~al.}, ``{Electron reconstruction in CMS}'',}
  \textit{ Eur. Phys. J. C} \textbf{ 49} (2007) 1099,
\href{http://dx.doi.org/10.1140/epjc/s10052-006-0175-5}{\doi{10.1140/epjc/s10052-006-0175-5}}.

\bibitem{CMSEGM}
\href {http://cdsweb.cern.ch/record/1299116} {{CMS Collaboration}, ``Electron
  Reconstruction and Identification at $\sqrt{s} = 7$ {TeV}'',} CMS Physics
  Analysis Summary CMS-PAS-EGM-10-004, 2010.

\bibitem{Blobel:2002pu}
\hrefCMSnoop {} {V.~Blobel, ``An unfolding method for high-energy physics
  experiments'',} in \textit{ Conference on Advanced Statistical Techniques in
  Particle Physics, Durham, England}, p.~258.
\newblock 2002.
\newblock
\href{http://www.arXiv.org/abs/0208022}{\texttt{ arXiv:0208022}}.
\newblock

\bibitem{Chatrchyan:2009sr}
\hrefCMSnoop {} {{ CMS} Collaboration, ``{Alignment of the CMS Silicon Tracker
  during Commissioning with Cosmic Rays}'',} \textit{ JINST} \textbf{ 05}
  (2010) T03009,
  \href{http://dx.doi.org/10.1088/1748-0221/5/03/T03009}{\doi{10.1088/1748-0221/5/03/T03009}}.

\bibitem{Alekhin:2011sk}
\hrefCMSnoop {} {S.~Alekhin {et~al.}, ``The PDF4LHC Working Group Interim
  Report'',} (2011).
\href{http://www.arXiv.org/abs/1101.0536}{\texttt{ arXiv:1101.0536}}.

\bibitem{Botje:2011sn}
M.~Botje\hrefCMSnoop {} { {et~al.}, ``The {PDF4LHC} {W}orking {G}roup Interim
  Recommendations'',} (2011).
\href{http://www.arXiv.org/abs/1101.0538}{\texttt{ arXiv:1101.0538}}.

\bibitem{PhysRevD.78.013004}
P.~M. Nadolsky\hrefCMSnoop {} { {et~al.}, ``Implications of {CTEQ} global
  analysis for collider observables'',} \textit{ Phys. Rev. D} \textbf{ 78}
  (2008) 013004,
  \href{http://dx.doi.org/10.1103/PhysRevD.78.013004}{\doi{10.1103/PhysRevD.78.013004}}.

\bibitem{Ball:2010de}
R.~D. Ball\hrefCMSnoop {} { {et~al.}, ``{A first unbiased global NLO
  determination of parton distributions and their uncertainties}'',} \textit{
  Nucl. Phys. B} \textbf{ 838} (2010) 136,
  \href{http://dx.doi.org/10.1016/j.nuclphysb.2010.05.008}{\doi{10.1016/j.nuclphysb.2010.05.008}}.

\bibitem{Martin:2009iq}
\hrefCMSnoop {} {A.~D. Martin, W.~J. Stirling, R.~S. Thorne, and G.~Watt,
  ``{Parton distributions for the LHC}'',} \textit{ Eur. Phys. J. C} \textbf{
  63} (2009) 189,
  \href{http://dx.doi.org/10.1140/epjc/s10052-009-1072-5}{\doi{10.1140/epjc/s10052-009-1072-5}}.

\end{thebibliography}\endgroup
